\RequirePackage{fix-cm}
\documentclass[twocolumn,epjc3]{svjour3}  
\smartqed  
 
\usepackage{dcolumn}  
\usepackage{bm}        
\usepackage{amssymb}
\usepackage{amsmath} 
\usepackage{epstopdf} 
\usepackage{amsfonts,float}
\usepackage{fancyhdr}
\usepackage{pifont}
\usepackage[hidelinks]{hyperref} 
\usepackage{bbm}
\usepackage{tensor}
\usepackage{mathrsfs}
\usepackage{xcolor}
\usepackage{physics}
\usepackage[sort&compress,numbers]{natbib}

\RequirePackage{graphicx}
		\newcommand{\bea}{\begin{eqnarray}} 
		\newcommand{\eea}{\end{eqnarray}}
		\newcommand{\beq}{\begin{equation}} 
		\newcommand{\eeq}{\end{equation}}
\journalname{Eur. Phys. J. C}
\begin{document}

\title{Thermal Casimir effect with general boundary conditions
}


\author{J. M. Mu{\~n}oz-Casta{\~n}eda\thanksref{e1,addr1}
        \and
        L. Santamar\'\i  a-Sanz\thanksref{e2,addr1} 
         \and
       M. Donaire\thanksref{e3,addr1} 
         \and
        M. Tello-Fraile\thanksref{e4,addr1} 
}

\thankstext{e1}{e-mail: jose.munoz.castaneda@uva.es}
\thankstext{e2}{e-mail: lucia.santamaria@uva.es}
\thankstext{e3}{e-mail: manuel.donaire@uva.es}
\thankstext{e4}{e-mail:marcos.tello@uva.es}


\institute{Departamento de F{\'\i}sica Te{\'o}rica, At{\'o}mica y {\'O}ptica, Valladolid University, Valladolid (SPAIN)\label{addr1}        
}

\date{Received: date / Accepted: date}

\maketitle

\begin{abstract}
In this paper we study the system of a scalar quantum field confined between two plane, isotropic, and homogeneous parallel plates at thermal equilibrium. We represent the plates by the most general lossless and frequency-independent boundary conditions that satisfy the conditions of isotropy and homogeneity and are compatible with the unitarity  of the quantum field theory. Under these conditions we compute the thermal correction to the quantum vacuum energy as a function of the temperature and the parameters encoding the boundary condition. The latter enables us to obtain similar results for the pressure between plates and the quantum thermal correction to the entropy. We find out that our system is thermodynamically stable for any boundary conditions, and we identify a critical temperature below which certain boundary conditions yield attractive, repulsive, and null Casimir forces.

\end{abstract}

\section{Introduction}

Since its theoretical prediction in 1948 \cite{Casimir} the Casimir effect has been extensively studied, both theoretically \cite{Milonni,Miltonbook,Buhmannbook,Bordagbook} and experimentally \cite{spa57,spa58,Lamoreaux,Decca}.  In its original formulation the Casimir force is a consequence of the interaction energy due to the coupling between the quantum vacuum fluctuations of the electromagnetic field with the charged current fluctuations of the plates \cite{Lifschitz,Lifschitz1}. For separation distances between plates much larger than any other length scale which determines the electric response of the plates, only the long-wavelength transverse modes of the electromagnetic field are relevant to the interaction, and they can be mimicked by the normal modes of a scalar field \cite{Miltonbook,Reynaudnotes,ReynaudJaks}.

Recently there has been renewed interest in the thermal Casimir effect motivated by its applications to the design of nano-electronic devices \cite{papers_on_chips,chips3,chips2,Sol-PRA2020}, the appearance of negative self-entropies in Casimir-like systems \cite{geye05-72-022111,lium19-100-081406,milt17-96-085007,li16-94-085010,bord18-98-085010,bord18-51-455001,bor-epjc}, technological applications, and cosmological pro-blems \cite{Weinberg}. In most of the cases the focus has been on the dependence of the Casimir effect at finite temperature with the geometry, and not much attention has been paid to the dependence on the physical properties of the boundaries appearing in the system.

The quantum vacuum energy at zero temperature of a massless scalar field confined between two parallel plates with general boundary conditions was studied in \cite{Asorey}, using the theory of selfadjoint extensions for the Laplace-Beltrami operator developed in \cite{Ibort}. The most remarkable result of Ref. \cite{Asorey} is the computation of the quantum vacuum energy for a scalar field confined between two homogeneous isotropic plates as a function over the space of those selfajoint extensions that are allowed in quantum field theory. As a consequence the authors were able to characterise those selfadjoint extensions that give rise to attractive, repulsive or null Casimir force between plates. In this article, we address the problem of the Casimir effect at finite temperature from a mathematical perspective. In particular, we compute the free energy, entropy and pressure, of a system described by a quantum scalar field in thermal equilibrium confined between two homogeneous parallel plates that are mimicked by the most general type of dispersionless and frequency-independent boundary conditions. Physically, the normal modes of the scalar field are a simplification of the transverse modes of the electromagnetic field, and the dispersionless and frequency-independent boundary conditions mimic the permitivities and permeabilities of  lossless plates which, for the range of wavelengths relevant to the problem, behave as effective constants. This simplification will enable us to understand the role of boundary conditions and its interplay with thermal quantum field fluctuations in the Casimir effect at finite temperature. In addition, it serves to extend the results of Refs. \cite{Ibort,Asorey} on a massless scalar field at zero temperature to the thermal environment. In addition, the calculation of the entropy will provide a clear understanding about the thermodynamical stability of the original Casimir system, but with sufficiently general boundary conditions. Finally, following the results of Ref. \cite{Asorey}, the calculation of the Casimir pressure between plates for general boundary conditions will allow us to distinguish between boundary conditions which produce repulsive, attractive or null Casimir force at finite temperature.

The article is organised as follows. In Section \ref{sec2} we make a compilation of basic formulas and previous results needed to obtain the main results of the paper. In Section \ref{sec3} we proceed to the calculation of the Helmholtz free energy and the entropy of a massless scalar field confined between two plates with general boundary conditions. Afterwards in Section \ref{sec4} we use the previous section's formulas to compute the Casimir quantum pressure between plates at finite temperature. We finalise with the conclusions in Section \ref{sec5}. In addition we have added an Appendix where we discuss the low and high temperature expansions for general boundary conditions that satisfy the requirements of isotropy and homogeneity.

Throughout the paper we will use natural units, $\hbar = c=k_B = 1$, being $\hbar$ the Planck constant, $c$ the speed of light, and $k_B$ the Boltzmann constant.

\section{Basic formulas}\label{sec2}

\subsection{Overview of scalar quantum fields with general boundary conditions.}

In this paper we will study a massless scalar field confined between two homogeneous and isotropic parallel plates. The susceptibility of the plates will be mimicked by the most general type of lossless and frequency-independent boundary conditions that are compatible with unitarity and satisfy the conditions of homogeneity and isotropy.  There exists a strong dependence of the quantum vacuum state and the  vacuum energy on the geometry of the physical space and the physical properties of the boundaries that interact with the quantum field, that are encoded in the  boundary conditions \cite{most-book,Bordagbook,Milonni,Miltonbook,BordagMohideen,Grahamjaffe,Mostepanenko1}. 
We consider a free massless complex scalar field $\phi$ confined in a domain $\Omega \in \mathbb{R}^3$ bounded by two parallel homogeneous and isotropic two-dimensional pla-tes orthogonal to the $x$-axis and placed at $x=0,L$, i. e., $\Omega=[0,L]\times\mathbb{R}^{2}$. In this situation the classical action for the massless scalar field that gives rise to local equations of motion is given by
\begin{equation}
S(\phi)=\frac{1}{2}\int_\Omega d^{3+1}x\, \partial^\mu\phi^*\partial_\mu\phi+\frac{1}{2} \int_{\partial \Omega} d^3 x \, \phi^* \partial_n \phi,
\end{equation} 
being $\partial_n$ the normal outgoing derivative. After a standard canonical second quantization the equation for the modes of the scalar quantum field is given by the non-relativistic Schr\"odinger eigenvalue problem,
\begin{equation}
-\Delta\phi_\omega(\boldsymbol{x})=\omega^2\phi_\omega(\boldsymbol{x}).
\end{equation}
Splitting the spatial coordinate as $\boldsymbol{x}=(x,\boldsymbol{y}_\parallel)$, with $\boldsymbol{y}_\parallel=(y,z)\in\mathbb{R}^{2}$ and $x\in[0,L]$, we can separate variables in the modes equation above by writing the modes of the quantum field as 
$\phi_\omega(\boldsymbol{x})=\psi_{\boldsymbol{k}_{\parallel}}(\boldsymbol{y}_\parallel)g_k(x)$. Under these assumptions the total Laplace operator and its spectrum of eigenvalues $\omega^2$ split as
\begin{equation*}
\Delta=\Delta_\parallel+\partial_x^2\Rightarrow\omega^2=\vec{k_\|}^2+k^2.
\end{equation*}
Assuming that the two plates are isotropic and homogeneous $\Delta_\parallel$ is nothing but the Laplace-Beltrami operator in $\mathbb{R}^{2}$, and its eigenvalues can be written as $\sqrt{\vec{k_\|}^2}$, with $\vec{k_\|}^2\in\mathbb{R}^2$. Hence the only nontrivial eigenvalue equation leftover is the one corresponding to the $OX_3$ axis,
\begin{equation}
-\frac{d^2}{d x^2}g_k(x)=k^2g_k(x), \quad x\in [0,L].
\end{equation}
The Laplace operator $\Delta$ over $\Omega=[0,L]\times\mathbb{R}^{2}$ is not essentially selfadjoint, so it admits an infinite set of selfadjoint extensions. In order to respect the unitarity principle of quantum field theory we must take into account only those selfadjoint extensions of the Laplace operator that give rise to non-negative selfadjoint operators for all $L \in (0, \infty)$. The set of selfadjoint extensions of $\Delta$ in $\Omega$ has been widely studied. From a physical point of view the most meaningful way to determine the set of selfadjoint extensions is given in Ref. \cite{Ibort}. Under our assumptions of homogeneity and isotropy of the plates the set of selfadjoint extensions of the Laplacian  $\Delta$ over $\Omega$ is in one-to-one correspondence with the set of selfadjoint extensions of the operator $-d^2/dx^2$ over $[0,L]$ which are given by the group $U(2)$ (see Ref. \cite{Asorey}). The domain\footnote{We denote by $H^2([0,L], \mathbb{C})$ the class $2$ Sobolev space.} $\mathcal{D}_U\subset H^2([0,L], \mathbb{C}) $ of field modes that defines the selfadjoint extension $\Delta_U$  is given in terms of the matrix $U \in U(2)$ (see \cite{Asorey, Munoz}) by
\begin{eqnarray}
&&\mathcal{D}_U= \lbrace \phi \in H^2([0,L], \mathbb{C}) / \varphi-i a\dot{\varphi}= U(\varphi+i a\dot{\varphi})\rbrace ,\label{bc}\\
&& \hspace{0.5cm}\varphi= \left( {\begin{array}{c}
   \phi(0) \\
   \phi(L) \\
  \end{array} } \right), \qquad \dot{\varphi}=\left( {\begin{array}{c}
   \partial_n \phi(0) \\
   \partial_n \phi(L) \\
  \end{array} } \right),\nonumber\\
  &&U(\alpha, \theta, \vec{n})= e^{i\alpha} [ \mathbbm{1}_2 \cos \theta + i (\vec{n} \cdot\boldsymbol{\sigma}) \sin \theta ],
\end{eqnarray}
where $\vec{n}=(n_1,n_2,n_3)$ is a unitary vector, $\boldsymbol{\sigma}$ is the vector of Pauli matrices, the angles $\alpha$, $\theta$ are such that $\alpha \in [0, \pi]$ and $\theta \in [-\pi/2, \pi/2]$, 
and $a$ is a fundamental length scale related to the electromagnetic response of the plates. Note that, except for the trivial choice $a=0$, in which case Eq.(\ref{bc}) reduces to Dirichlet's boundary conditions, Eq.(\ref{bc}) leaves us with two independent length scales at our disposal, in addition to the temperature.  From a physical perspective, considering $\phi$ as the effective field of transverse electromagnetic modes, and noting that $a$ relates field values and field derivatives, the length $a$ can be associated to the relationship between the electric and magnetic response of the plates. 

The space of boundary conditions $\mathcal{M}_F$ that give rise to non-negative selfadjoint extensions  $\Delta_U$ of the Laplacian operator is\footnote{Since there is a one-to-one correspondence between unitary matrices $U$ determining a boundary condition, and selfadjoint extensions $\Delta_U$ we will not distinguish between the unitary matrices and their corresponding selfadjoint extensions throughout the paper.} 
\begin{equation}
\mathcal{M}_F \equiv \lbrace U(\alpha, \theta, \boldsymbol{n}) \in U(2) / 0 \leq \alpha\pm \theta \leq \pi \rbrace .
\end{equation}
We can characterize the non-zero part of the spectrum for any selfadjoint extension   $\Delta_U \in \mathcal{M}_F $  including multiplicities of eigenvalues throughout the secular equation obtained in \cite{Asorey},
\begin{eqnarray}
h_{U}(k)&=&\sin{(kL)}[(k^{2}a^{2}-1)\cos{\theta}+(k^{2}a^{2}+1)\cos{\alpha}]\nonumber\\
&-&2ka\cos{(kL)}\sin{\alpha}-2ka\:n_{1}\sin{\theta},\label{fU}
\end{eqnarray}
where, by assumption (isotropy and homogeneity of the plates), the boundary condition parameters do not depend on $k$ and are uniform on the plates. In addition, since they are temperature independent, Eq.(\ref{fU}) is equivalent to the spectrum of normal modes obtained in Refs.\cite{Asorey,Munoz}. Note that since $f_{U}$ contains terms in different powers of the dimensionless quantities $kL$ and $ka$, the spectral function varies with respect to $a$ and $L$ in an independent manner. Hereafter, in order to simplify matters we  will disregard the trivial case $a=0$ and consider without loss of generality $a=1$, unless stated otherwise. Thus, the separation length $L$ and the inverse of the temperature $T^{-1}$ will be expressed in units of $a$ in most of the reminder of this paper.

 
The vacuum energy is given by the sum of the eigenvalues of $\sqrt{-\Delta_U}$, i. e.,
$
E_0= \textrm{tr} \sqrt{-\Delta_U}
$. 
This sum is ultraviolet divergent due to the contributions of the energy density of the field theory in the bulk and the surface energy density associated to the plates. Nonetheless, there are finite volume corrections to the vacuum energy that give rise to a finite neat Casimir effect. These divergencies can be subtracted to obtain a finite result. Following \cite{Asorey,bor-mpla19} we can write the zero temperature finite Casimir energy per unit area of the plates in three dimensions as
\beq \label{e0fin}
\frac{E^{(3)}_U}{A}=\frac{1}{6\pi^2}\int_0^\infty \!\!\!\!\!dk\,k^3\!\!\left[ L-\partial_k\log\left( \frac{h_U(ik,L)}{h_U^\infty(ik)}\right) \right]\!,
\eeq
where $A$ is the area of the plates and $h_U^\infty(ik)$ is the dominant asymptotic term of $h_U(ik)$ as $L\to\infty$, given by \cite{bor-mpla19}
\beq
\lim_{L_0\to\infty}\frac{h_U(ik,L_0)}{e^{k L_0}}\equiv h_U^\infty(ik).
\eeq
Defining the polynomial $c_U(z)\equiv z^2-z\, {\rm tr}(U)+\det(U)$ it is easy to see that
\beq
 h_U^\infty(ik)=\frac{1}{2}(k-i)^2c_U\left(-\frac{k+i}{k-i}\right).
\eeq

\paragraph{Comment on the electromagnetic field.} It is common in the literature concerning electromagnetic Casimir forces to split the electromagnetic field modes into {\it transverse electric} (TE) and {\it transverse magnetic} (TM) polarizations. Effectively this splitting enables to treat the TE and TM modes as independent scalar fields, specially concerning the boundary conditions in a parallel plates setup. For an ideal conductor, i. e., one for which the permittivity $\varepsilon$ tends to infinity,  the TE and the TM modes  will satisfy Dirichlet and Neumann boundary conditions respectively. In the case where the permeability $\mu$ tends to infinity it is the TE-modes the ones that satisfy Neumann boundary condition meanwhile the TM-modes verify Dirichlet boundary condition. For intermediate situations where the plates  have non-inifite permittivity or permeability the transverse electromagnetic modes verify Robin boundary conditions. In all this situations the plates are mimicked by boundary conditions with the formalism presented above. Specifically, Dirichlet boundary condition for both plates is obtained when one takes $U=-\mathbbm{1}_2$,  Neumann boundary condition arises for $U=\mathbbm{1}_2$ and Robin boundary condition for identical plates (both plates have the same Robin parameter) can be obtained when $U=e^{i\alpha}\mathbbm{1}_2$ with $\alpha\in[0,\pi]$. In all these cases the plates are mimicked by boundary conditions with constant parameters. Generally, in realistic materials the response of the plates depend on the frequency as well as on the parallel components of the momentum of the field modes, $\vec{k_\|}$. If we restrict ourselves to frequency-indepen-dent boundary conditions, the formalism presented here can be extended by letting the parameters of the boundary conditions depend on $\vec{k_\|}$, i. e., $$U=U(\alpha(\vec{k_\|}),\theta(\vec{k_\|}),\boldsymbol{n}(\vec{k_\|})).$$ In this generalised situation the boundary condition in Eq. \eqref{bc} remains valid and one would have to account for an spectrum of normal momenta that is not independent of the parallel momenta. Examples of this situations in Casimir setups can be found in Refs.  \cite{Vassilevich-PRB-2011, Mostepanenko-PRA-2012}, where the effective couplings between the plates and the electromagnetic field in between arise after the integration of the electron Dirac field in the plates\footnote{The response of the plates is due to the 1-loop effective action of the fields in the plates that are coupled to the electromagnetic filed. In this situation in general the effective couplings that mimmic the plates are frequency dependent.}. Specifically in Eq. (17) from Ref. \cite{Vassilevich-PRB-2011} it is shown how the electromagnetic 4-potential $A_\mu$ satisfies a boundary condition that defines the Dirac-$\delta$ with a coupling given by the polarization tensor emerging after integrating over the Dirac fields over graphene plates. The Dirac-$\delta$ boundary conditions can be easily written in the form of Eq. \eqref{bc} as was shown in Ref. \cite{Guilarte-PRD-2015}, but since the polarization tensor shown in Ref. \cite{Vassilevich-PRB-2011} depends on the parallel momenta $\vec{k_\|}$ the parameters of the corresponding matrix $U$ will depend on $\vec{k_\|}$. For simplicity in this paper we will only consider cases in which the matrix $U$ does not depend on the parallel momenta 
$\vec{k_\|}$ (isotropy requirement) since it enables us to infer which are the effects of the boundary conditions with enough generality.

\paragraph{Boundary conditions and topology change.} The formalism developed in Refs. \cite{Ibort,Asorey} for boundary conditions enables the implementation of topology changes in the physical space. This can be seen easily by noticing that there are one-parameter families of boundary conditions in ${\cal M}_F$ that interpolate smoothly between a system with two plates and a system where these two plates are identified to give rise to a cylinder. The simplest example showing this situation is the one-parameter family of boundary conditions defined by the unitary operator
\begin{equation}\label{utheta}
U(\theta)=-e^{i\theta}(\mathbbm{1}_2\cos(\theta)+i\sigma_1\cos(\theta)),
\end{equation}
for $ \theta\in[-\pi/2,0]$. On the one hand, from Eq. \eqref{utheta} $U(\theta=0)=-\mathbb{I}$ which corresponds to Dirichlet boundary condition, where one has two identical plates that mimic the interaction of an electromagnetic TE mode with two perfectly conducting identical parallel plates. On the other hand, $U(\theta=-\pi/2)=\sigma_1$ defines the well known periodic boundary condition where the interval $[0,L]$ is identified with a circle with length $L$ and we can not speak about identical plates. This example illustrates how variations in the parameters of the boundary conditions involve topology changes (see Ref. \cite{Ibort} for a more detailed discussion).

\subsection{Free energy and thermodynamics.}

When considering thermal excitations of an ensemble of particles, the statistical behaviour of the ensemble is characterised by a temperature $T$ and a probability distribution once the equilibrium is reached. In our case, a quantum scalar field between plates is nothing but an infinite collection of harmonic oscillators that do not interact with each other. The system is characterised by the grand canonical partition function, ${\cal Z}(T)$, which will be computed in the next section.
The Helmholtz free energy of the system in thermal equilibrium at a temperature $T$ is given in terms of the partition function as
\begin{equation}\label{free_energy}
\mathcal{F} = -T\ln {\cal Z}.
\end{equation}
Once the free energy is known, other thermodynamic quantities can be obtained easily. In particular, we will focus our attention in the entropy ($S$)
\begin{equation}\label{entropy}
S = -\frac{\partial\mathcal{F}}{\partial T},
\end{equation}
and the force between plates per unit of area of the plates, i. e., the pressure $P$,
\begin{equation}\label{e15}
P=-\frac{1}{A}\left(\frac{\partial\mathcal{F}}{\partial L}\right)_T.
\end{equation}
The main aim of this work is to calculate the  thermal correction $\Delta_T\mathcal{F}$ to the zero-temperature vacuum energy $E_U^{(3)}$, the entropy, and the quantum vacuum pressure of the system for arbitrary temperature and arbitrary boundary conditions fulfilling the requirements of homogeneity and isotropy.

\section{Free energy and entropy at finite temperature }\label{sec3}
 

The fact that the unitary matrix ${\bf U}\in U(2)$ that defines the boundary condition in Eq. \eqref{bc} does not depend on $k$, together with the unitarity of the quantum field theory ensured by the non-negativity of $-\Delta_U$, enables us to simplify the expression of the free energy.
Indeed, in our case the Hamiltonian of the quantum field theory can be written as a formal summation over the modes of the quantum field as
\begin{equation}
{\cal H}=\sum_{\omega^2\in\sigma(\Delta_U)} \omega\left(\hat N_\omega+\frac{1}{2}\right),
\end{equation}
where $\sigma(\Delta_U)$ is the spectrum of the corresponding selfadjoint extension $\Delta_U$. Therefore we can simply write
\begin{equation}
e^{-{\cal H}/T}=\prod_{\omega^2\in\sigma(\Delta_U)}\exp\left( -\frac{ \omega}{T}\left(\hat N_\omega+\frac{1}{2}\right) \right).
\end{equation}
 Thus, we can treat the system as an ensemble of non-interacting harmonic oscillators with energy levels
\begin{eqnarray}
&&E_n= \left(\frac{1}{2}+n \right)  \omega , \qquad  n\in 0 \cup \lbrace\mathbb{N}\rbrace,
\end{eqnarray}
where $\omega^2$ is an eigenvalue of the selfadjoint extension of $\Delta_U$ with boundary condition given by a certain matrix $U\in {\cal M}_F$, i. e., the non-zero eigenvalues $\{\omega^2\}$ are given by the zeros of $h_U$ in Eq.(\ref{fU}).
The fact that the ensemble of harmonic oscillators do not interact enables to write the partition function of the quantum field theory as an infinite product of harmonic oscillator canonical partition functions, one for each frequency $\omega$ ($\omega^2\in\sigma(\Delta_U)$),
\beq\label{eq20}
{\cal Z}=\prod_{\omega^2\in\sigma(\Delta_U)}Z_{os}(T;\omega).
\eeq
It is a well known result that the canonical partition function for a single harmonic oscillator of frequency $\omega$ can be written as \cite{reichl}
\begin{eqnarray}
Z_{os}(T)&=& \sum_{n=0}^\infty e^{- E_n/T}= 
\frac{e^{- \omega/2T}}{1-e^{-\omega/T}},
\end{eqnarray}
and the corresponding free energy is
\begin{equation}
\mathcal{F}_{os}(T)= -T \ln Z_{os} = \frac{\omega }{2} +T \,\textrm{ln} (1- e^{-\omega/T}).
\end{equation}
Hence from this expression and using Eqs. \eqref{free_energy} and \eqref{eq20} after some straightforward manipulations we obtain for the total Helmholtz free energy the well known result
\begin{equation}
\mathcal{F}(T)= \sum_{\omega^2  \in \, \sigma (\Delta_U)}\left[ \frac{\omega }{2} + T  \,\textrm{ln} (1- e^{-\omega/T})\right].
\end{equation}
From the previous equation we can split the free energy \eqref{free_energy} in two parts \cite{Bordagbook},
\begin{equation}\label{FE_total}
\mathcal{F} = \left.\mathcal{F}\right\vert_{T=0}+ \Delta_T\mathcal{F}.
\end{equation}
The first one corresponds to the quantum vacuum energy at zero temperature,
\begin{equation}
E_0^{vac}\equiv \left.\mathcal{F}\right\vert_{T=0}=\frac{1 }{2}\sum_{\omega^2  \in \, \sigma (\Delta_U)}\!\!\!\!\omega ,
\end{equation}
which carries all the divergences; thus, it must be renormalized \cite{Bordagbook}. The divergences and regularization methods for $E_0^{vac}$ have been largely studied (see e.g. Refs. \cite{Bordagbook,Milonni,Miltonbook}), and in our case the finite quantum vacuum energy will be given by \eqref{e0fin} following Refs. \cite{Asorey,bor-mpla19}.
The second part of \eqref{FE_total} is the temperature dependent part of the free energy that is free of divergencies and can be written as
\begin{eqnarray}\label{eq14}
\Delta_T \mathcal{F}&=& \sum_{\omega^2  \in \, \sigma (\Delta_U)} B(\omega, T) \\
B(\omega,T)&=& T\ln \left[ 1- \textrm{exp}\left(-\frac{\omega}{ T}\right) \right].\label{Bfactors}
\end{eqnarray}
The summation over the field modes $\omega$ can be separated into the summation over the parallel momenta, which is an integration over $ \vec{k_\|}$, and the discrete summation over the transverse momenta $k$.

\subsection{Summation over the parallel momenta}

Starting from Eq. \eqref{eq14} and taking into account that the frequencies of the field modes are given by $\omega=\sqrt{\vec{k_\|}^2+k^2}$, with $\vec{k_\|}$ being the two-dimensional parallel momenta and $k$ the discrete orthogonal momenta (which can be obtained from the non-null zeroes of the spectral function, i.e., $Z^*(h_U)$). Hence the summation over the whole spectrum $\sigma (\Delta_U)$ when $\Delta_U$ does not have zero modes transforms into
\begin{equation}
\bigtriangleup_T \mathcal{F}= \!\!\sum_{\sigma (\Delta_U)} \!\!B(\omega, T)= {\color{black}A} \!\!\int_{\mathbb{R}^{2}}  \!\!\frac{d^{2} \vec{k_\|}}{{\color{black}(2\pi)^2}} \!\! \! \sum_{k\in Z^*(h_U) }\!\!  \!\!B(\omega, T),
\end{equation}
where $A$ is the area of the plates. The integration over the parallel momenta can be commuted with the summation over the discrete transverse momenta. Doing so the integration over the parallel momenta reads
\begin{eqnarray}
&&I_3(k,T)=T\int_{\mathbb{R}^{2}} \frac{d^{2} \vec{k_\|}}{{\color{black}(2\pi)^2}}\,   \log \left( 1-e^{-\frac{\sqrt{k_\|^2+k^2}}{ T}}\right)=\nonumber\\
&&\frac{T k}{{\color{black} 2\pi}} \int_0^\infty dk_\| \, \frac{k_{\|}}{k} \,   \log \left( 1-e^{-\frac{k}{T}\sqrt{k_\|^2/k^2+1}}\right).\label{i3-def}
\end{eqnarray}
This integration can be performed analytically using {\it Mathematica}, to obtain
\begin{equation}\label{eq26}
I_3(k,T)=-  \frac{T^3}{\color{black} 2\pi} \left(\frac{k}{T}\text{Li}_2\left(e^{-\frac{k}{T}}\right)+\text{Li}_3\left(e^{-\frac{k}{T}}\right)\right),
\end{equation}
where $\text{Li}_s(z)$ denotes the polylogarithmic function of order $s$ \cite{nist-book}.

We are just left the summation over transverse momenta $k$. At this point we need to distinguish between those boundary conditions that give rise to selfadjoint extensions that do not have zero modes, and those that do. The subset ${\cal M}_F^{(0)}\in {\cal M}_F$ of selfadjoint extensions that admit zero modes was characterised in Ref. \cite{Munoz}. Furthermore in Ref. \cite{Munoz} it was demonstrated that any $\Delta_U\in {\cal M}_F^{(0)}$ only admits one constant zero-mode.

\subsection{The case with no zero-modes: $U\in {\cal M}_F-{\cal M}_F^{(0)}$}

\begin{figure}[H]
\centering
\includegraphics[scale=0.55]{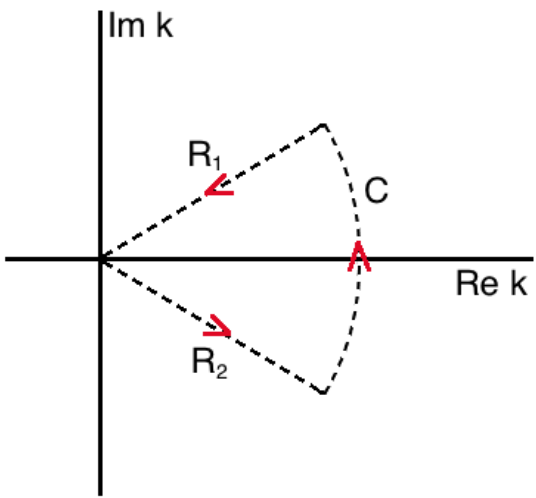}
\caption{\small Complex contour $\Gamma$ that encloses all the zeroes of $f_U(k)$ as $R\to \infty$. In this figure, $R_1:\,z=\xi e^{i\gamma}$ with $\xi\in[R,0]$; $R_2:\,z=\xi e^{-i\gamma}$ with $\xi\in[0,R]$;  $C:\, z=R e^{i\mu}$ with $\mu\in[-\gamma,\gamma]$. For the contour $\Gamma$, $R>0$ and $0<\gamma<\pi/2$ constants.}\label{fig1}
\end{figure}

{\color{black} As stated above the summation over transverse momenta is equivalent to summing over the zeros of $h_U(k)$ different from $k=0$. As was explained in Ref. \cite{Munoz}, when $U\in{\cal M}_F-{\cal M}^{(0)}_F$ the spectral function $h_U(k)$ from Eq. \eqref{fU} needs to be replaced with
\begin{equation}\label{fug}
f_U(k)\equiv k^{-1}h_U(k),
\end{equation}
in order to be able to write the summation over the discrete transverse momenta as a contour integral \cite{Klaus,piroz-prd57} avoiding the possible problems at $k=0$.}
Hence the final formula for the temperature dependent part of the free energy when $\Delta_U\in{\cal M}_F-{\cal M}^{(0)}_F$ is
\begin{eqnarray}\label{eq28}
&&\Delta_T \mathcal{F}= {\color{black}A}\sum_{k\in Z(f_U) }  I_3(k,T) ,
\end{eqnarray}
being $Z(f_U)$ the set of zeros of $f_U(k)$.
The summation in Eq. \eqref{eq28} can be written down by using a complex contour integral as
\begin{eqnarray}\label{eq28}
 &&\Delta_T \mathcal{F}=\lim_{R\to\infty} {\color{black} A}\oint_\Gamma \frac{dk}{2\pi i} I_3(k,T)\,  \partial_k \log f_U(k),
\end{eqnarray}
where $\Gamma$ is the contour shown in  Fig. \ref{fig1}, which encloses all the zeroes of $f_U(k)$ when $R\to\infty$. 
The  integral \eqref{eq28} is well defined because $f_U(k)$ is a holomorphic function on $k$.  
When $R\to\infty$ in the contour of Fig. 1, the integration over the circumference arc $C$  goes to zero. Hence integrating over the whole contour in Fig. 1, and taking the limit $R\to \infty$, Eq. \eqref{eq28} reduces to the integration over the two straight lines  $z=\xi e^{\pm i\gamma}$ with being $\gamma$ a constant angle and $\xi\in [0,\infty)$,
\begin{eqnarray}\label{eq18}
&&\bigtriangleup_T \mathcal{F}= {\color{black} A}\int_0^\infty \frac{d\xi}{2\pi i } \left[-I_3(\xi e^{i\gamma},T) \partial_\xi \log f_U(\xi e^{i\gamma}) +\right.\nonumber\\
&&\left. \hspace{2cm} I_3(\xi e^{-i\gamma},T) \partial_\xi \log f_U(\xi e^{-i\gamma})\right].
\end{eqnarray}
The residue theorem ensures that the result of this integration does not depend on the angle $\gamma$ taken in the contour. This is so because all the zeros of $f_U(z)$ lie in $\mathbb{R^+}$ since $\Delta_U\in{\cal M}_F-{\cal M}_F^{(0)}$ has no zero modes and it is a definite positive selfadjoint operator \cite{Asorey}. 
\begin{figure*}
\centering
    \includegraphics[width=1\textwidth]{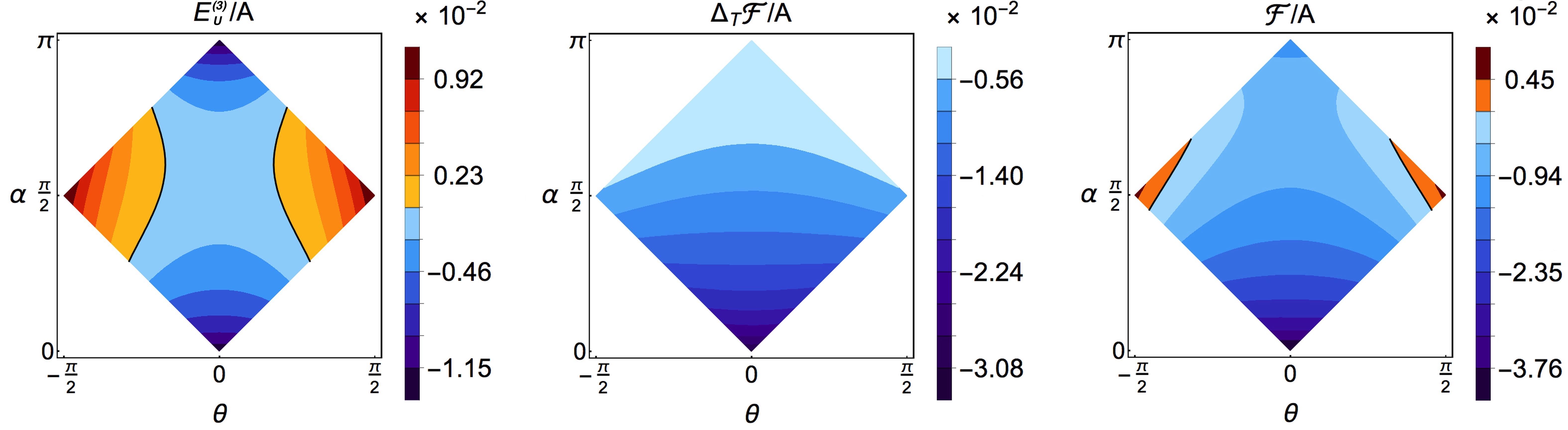}
    \caption{Quantum vacuum energy per unit area at $T=0$ (left), thermal correction $\Delta_T{\cal F}/A$ (center) and total Helmholtz free energy per unit area (right) as functions of the parameters $\alpha$ and $\theta$. In these plots, $ T=0.55, n_1=0$ and $L=1$.}
     \label{fig2}
\end{figure*}
\begin{figure*}
\centering
    \includegraphics[width=1\textwidth]{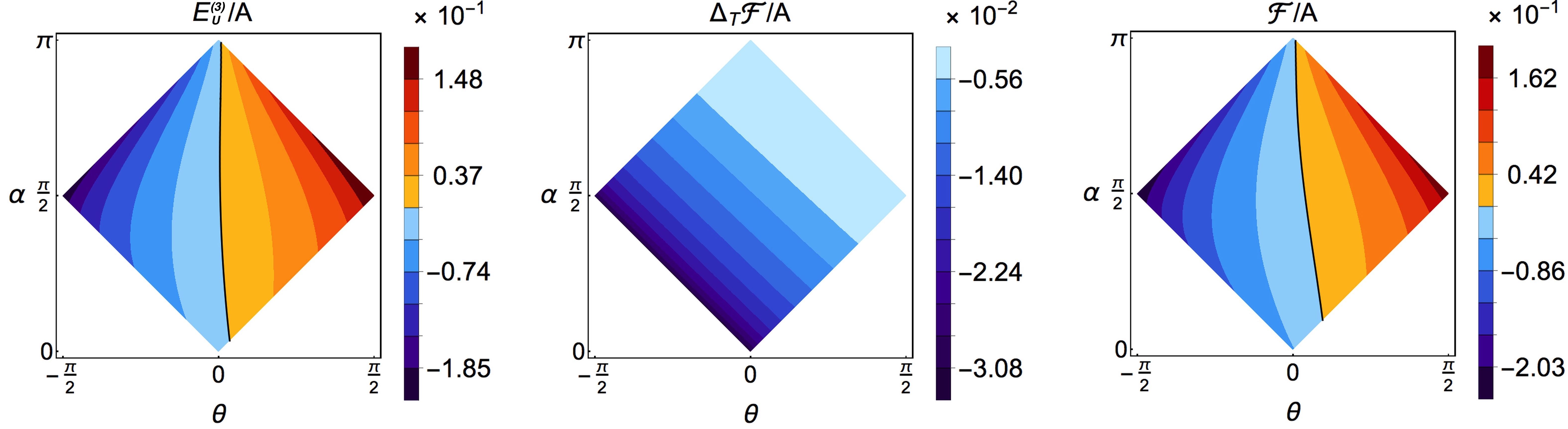}
    \centering
     \caption{Quantum vacuum energy per unit area at $T=0$ (left), thermal correction $\Delta_T{\cal F}/A$ (center) and total Helmholtz free energy per unit area (right) as functions of the parameters $\alpha$ and $\theta$. In these plots, $T=0.55, n_1=1$ and $L=1$.}
     \label{fig3}
\end{figure*}
\begin{figure*}
\centering
    \includegraphics[width=1\textwidth]{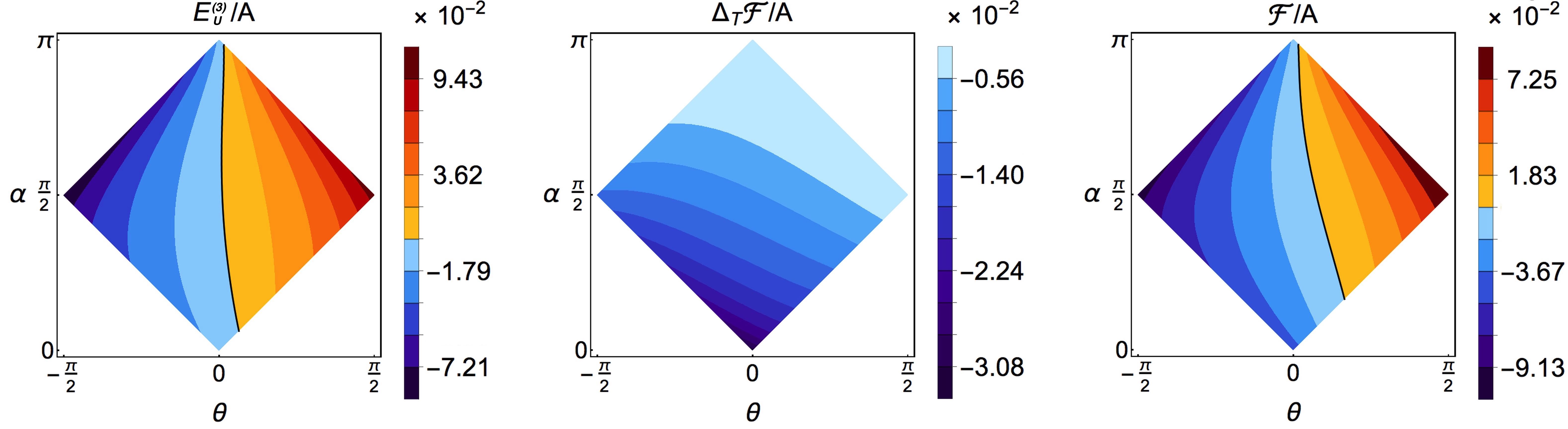}
     \caption{Quantum vacuum energy per unit area at $T=0$ (left), thermal correction $\Delta_T{\cal F}/A$ (center) and total Helmholtz free energy per unit area (right) as functions of the parameters $\alpha$ and $\theta$. In these plots, $T=0.55, n_1=0.5$ and $L=1$.}
     \label{fig4}
\end{figure*}
\begin{figure*}
\centering
    \includegraphics[width=1\textwidth]{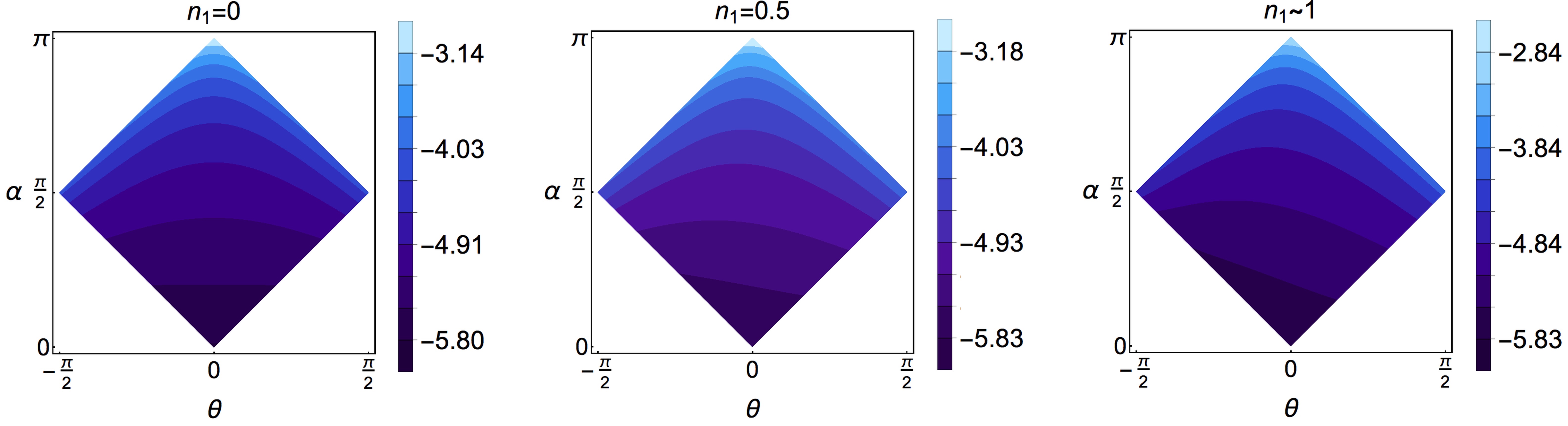}
     \caption{$\Delta_T{\cal F}/A$ for hight temperature (T=2.5) as function of the parameters $\alpha$, $\theta$ and $n_1$. In these plots we have fixed $L=1$.}
     \label{fig5}
\end{figure*}
It is of note that when we take $\gamma=\pi/2$  and Cauchy Principal Values are considered, we obtain the well known Matsubara formula after integrating by parts in $\xi$ as shown in Refs \cite{Lifschitz,Drudeplasma1,Bordagbook,BuhmannScheel,bor-epjc,bor-jpa53}.


The integral \eqref{eq18} can be evaluated numerically with \textit{Mathematica} for any finite temperature $T$ and any unitary matrix $U\in{\cal M}_F-{\cal M}_F^{(0)}$ to obtain $\Delta_T{\cal F}(U)$. In addition, formula \eqref{e0fin} from Refs. \cite{Asorey,bor-mpla19} enables to obtain the zero temperature energy which, together with $\Delta_T{\cal F}(U)$ give the total Helmholtz free energy as a function of the temperature $T$ and the parameters of the general boundary condition.

\paragraph{Low temperature behaviour of the Helmholtz energy.}In Figs. \ref{fig2}-\ref{fig4} we can observe the numerical results for the free energy at low temperatures ($T=0.55$). In each figure we can see the quantum vacuum energy at $T=0$ (left plots) computed with formula \eqref{e0fin}, the thermal correction $\Delta_T{\cal F}$ (central plots) and the total free energy ${\cal F}$ (right plots) as functions of the parameters $\{\alpha,\theta,n_1\}$ defining the boundary condition. 
It can be seen that although the thermal correction $\Delta_T{\cal F}$ is definite negative for any boundary condition, the total free energy can be positive, negative, or zero, as for the case of the quantum vacuum energy at zero temperature \cite{Asorey}. On the contrary, unlike it happens for $T=0$ where the vacuum energy behaves with the distance between plates as
$E_U^{(3)}/A\sim L^{-3}$,
for any $T>0$, positive or negative total Helmholtz free energy does not ensure attractive or repulsive thermal Casimir force. This is discussed in detail in the next section.

\paragraph{High temperature behaviour of the Helmholtz energy.} The thermal correction $\Delta_T{\cal F}$ is a negative and monotonically decreasing function of $T$, as can be seen in Fig. \ref{fig5}. Hence, as the temperature grows it dominates the total Hemholtz free energy. In addition, if we compare the plots for $E_U^{(3)}$ in Figs. \ref{fig2}-\ref{fig4} with the plots in Fig. \ref{fig5} it is straightforward to see that for $TL\gg1$,
 \begin{equation}
 \vert \Delta_T{\cal F}\vert\gg\vert E_U^{(3)}\vert.
 \end{equation}
 Therefore at high temperature the total Helmholtz free energy is always negative. For further details about the high temperature behaviour see the Appendix.

\begin{figure}
\centering
    \includegraphics[width=.40\textwidth]{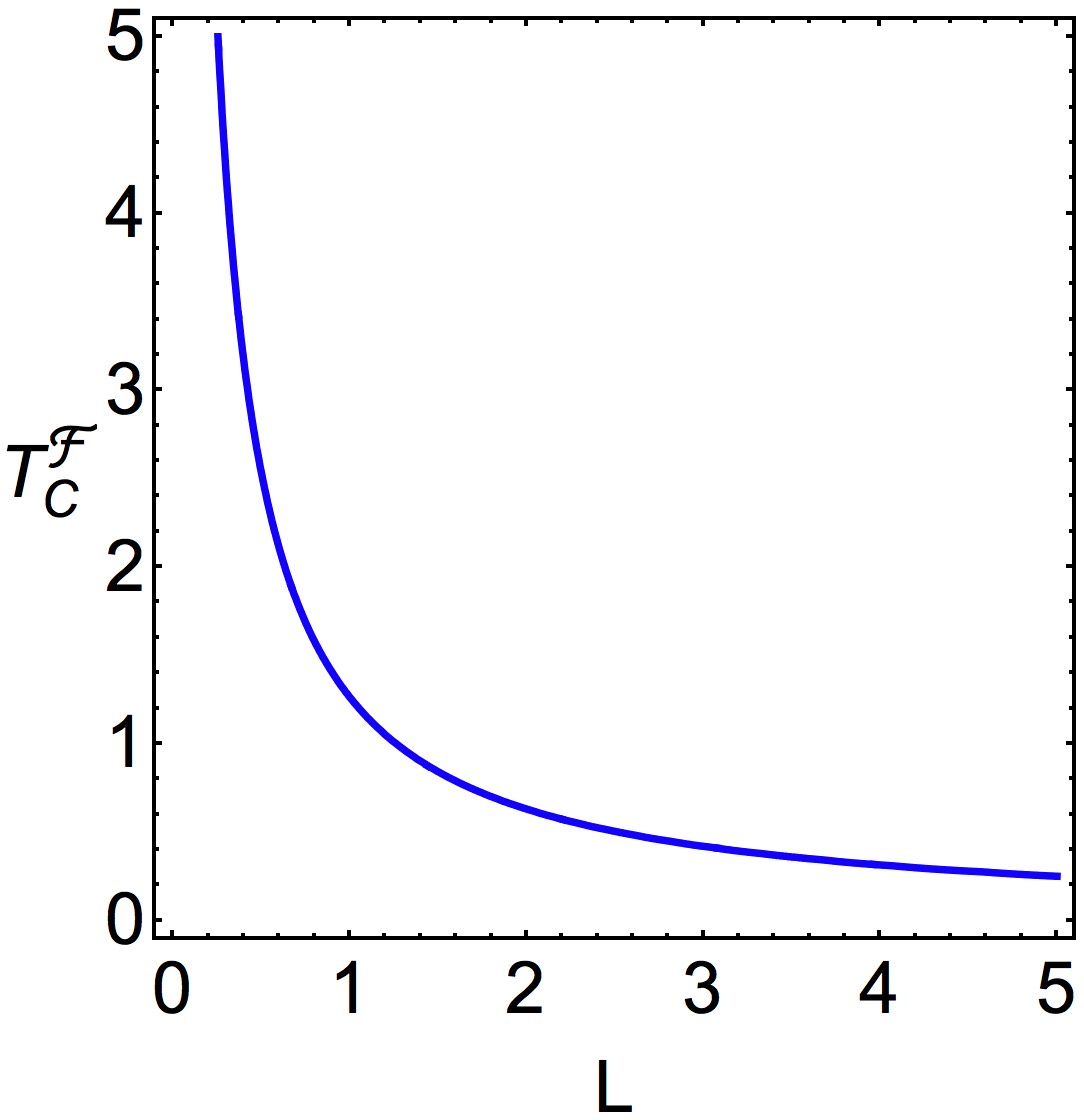}
     \caption{$T_c^{\cal F}$ as a function of the distance between plates $L$.}
     \label{fig6}
\end{figure}

\paragraph{On the critical temperature $T_c^{\cal F}$.} From the numerical results discussed above, we infer that for a fixed length $L$ there should be a critical temperature $T_c^{\cal F}$ such that for any $T>T_c^{\cal F}$ there are no boundary conditions giving rise to positive total Helmholtz free energy. On the other hand, whenever $T<T_c^{\cal F}$ total Helmholtz energy ${\cal F}$ will not have a defined sign, i. e., it can be positive, negative or zero. From  Ref. \cite{Asorey} we know that the boundary condition for which the quantum vacuum energy at $T=0$ reaches its maximum is given by anti-periodic boundary conditions,
 \begin{equation}
U_{ap}=\left(
\begin{array}{cc}
 0 & -1 \\
 -1 & 0 \\
\end{array}
\right).
\end{equation}
Moreover, the quantum vacuum energy for anti-periodic boundary conditions can be obtained analytically,
\begin{equation}
E_{ap}=A\frac{7 \pi ^2}{360 L^3}.
\end{equation}
Since $\Delta_T{\cal F}$ is a monotonically decreasing function of $T$, we can ensure that the situation in which the possibility of having positive total Helmholtz free energy at a given $T$ disappears occurs when 
\begin{equation}\label{full-neg-cond}
\vert \Delta_T{\cal F}(U_{ap})\vert=E_{ap}.
\end{equation}
This equality yields an equation in $T$ and $L$ that enables us to obtain $T_c^{\cal F}$ numerically. In Fig. \ref{fig6} it is shown $T_c^{\cal F}$ as a function of the length $L$. It is of note that the critical temperature $T_c^{\cal F}$ does not separate the regimes in which the quantum vacuum force is fully repulsive and the case in which it can be repulsive, attractive or zero.

\begin{figure*}
\centering
    \includegraphics[width=0.7\textwidth]{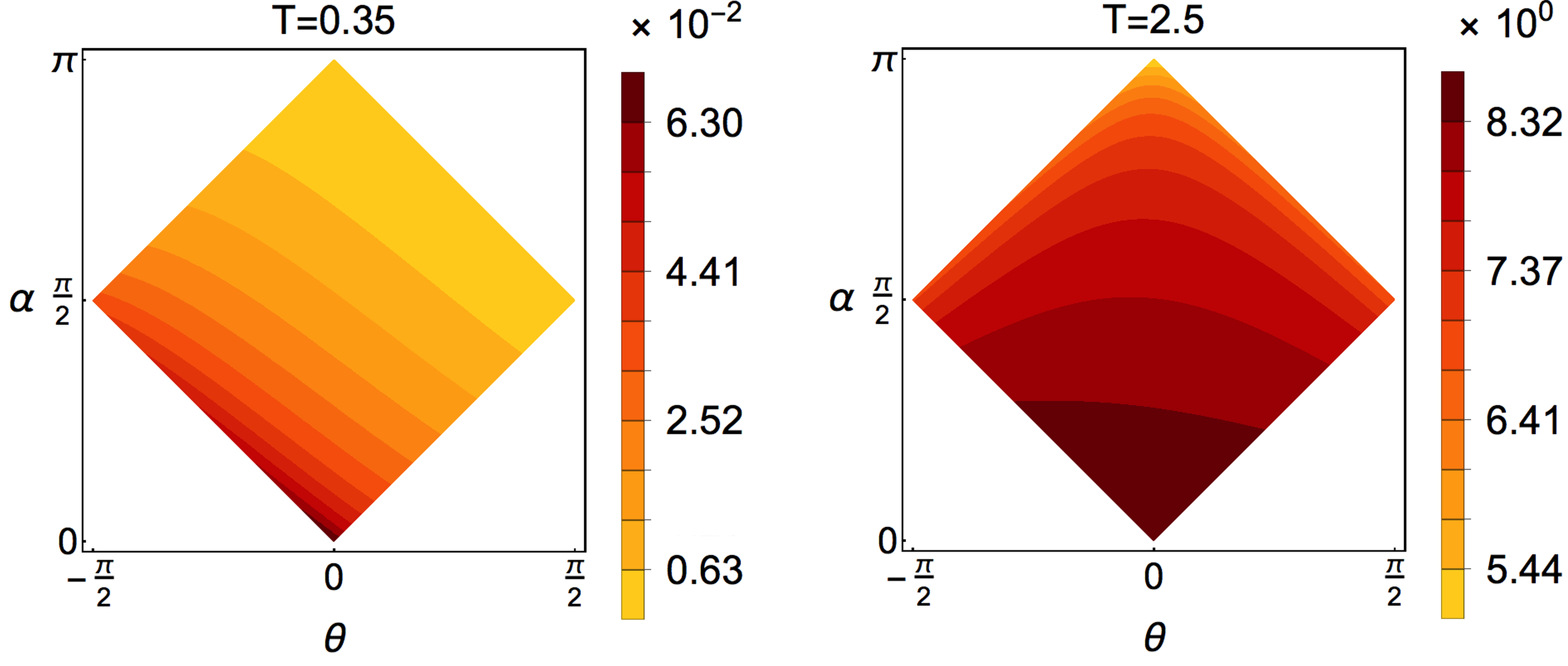}
     \caption{Entropy per unit area for low (left) and high (right) temperatures with $n_1=0.75$. In these plots $L=1$.}
     \label{fig7}
\end{figure*}
\begin{figure*}
\centering
    \includegraphics[width=0.7\textwidth]{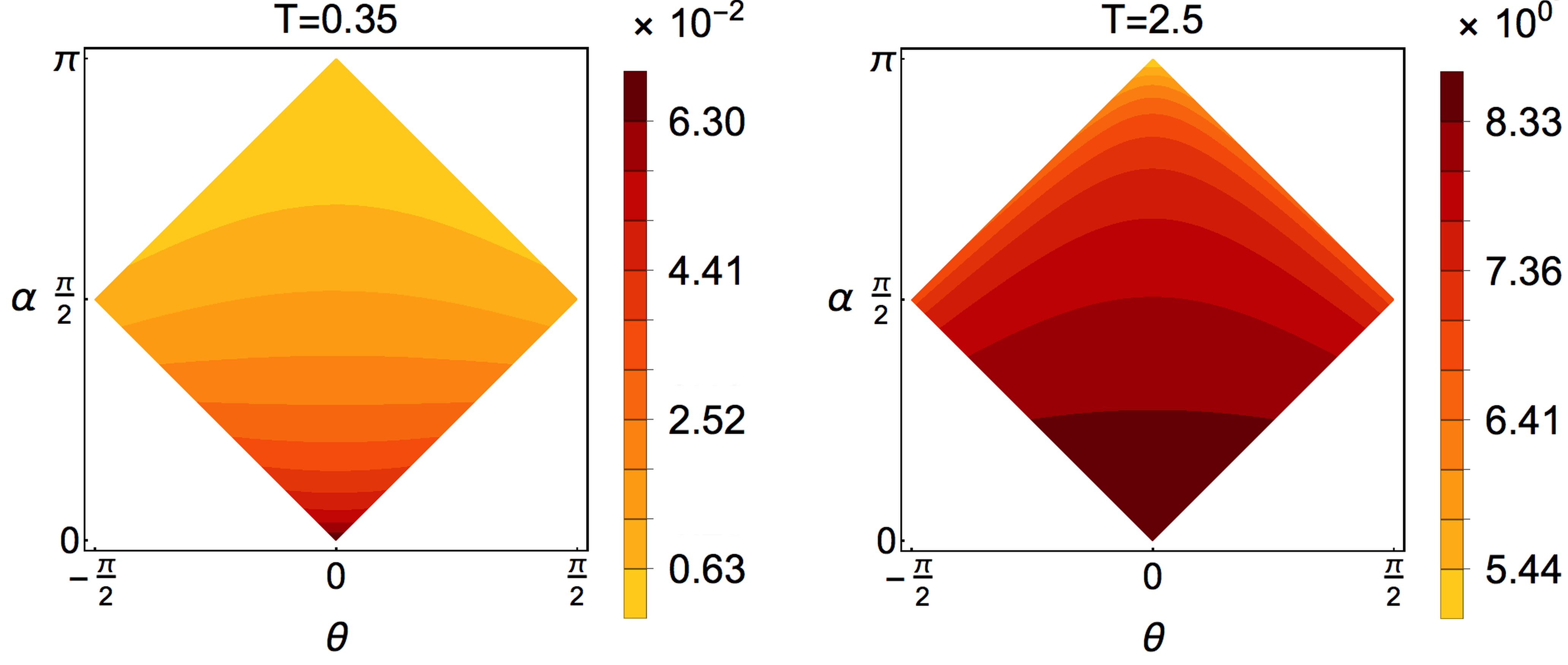}
     \caption{Entropy per unit area for low (left) and high (right) temperatures with $n_1=0$. In these plots $L=1$.}
     \label{fig8}
\end{figure*}
\paragraph{The entropy.} The entropy arising from Casimir self-energies has been a field of intensive activity in recent years since it was noticed in Ref. \cite{geye05-72-022111} that the quantum vacuum energy at finite temperature can give rise to negative corrections to the entropy. The existence of negative entropy is interpreted as a hint of possible instabilities in Casimir-like systems \cite{thir70-235-339}. Therefore, the calculation of the entropy for the system we are studying is of great interest to infer if there are boundary conditions that can generate negative entropy corrections. Making use of formula \eqref{entropy} we can compute numerically the entropy as a function over the space of boundary conditions for any arbitrary temperature (see Figs. \ref{fig7}-\ref{fig8}). As can be seen the entropy is positive definite for any set of parameters $\{\alpha,\theta,n_1\}$ and is a monotonically increasing function of $T$. This ensures that there is no boundary condition giving rise to negative entropy corrections. Therefore all the boundary conditions are thermodynamically stable. Moreover for any $T>0$ the maximum entropy is reached for Neumann boundary condition ($\alpha=\theta=0$) and minimum entropy occurs for Dirichlet boundary condition ($\alpha=2\pi, \,\theta=0$).

\subsection{Quantum field theories with zero modes: $\Delta_U\in{\cal M}_F^{(0)}$}
 The boundary conditions in ${\cal M}_F$ that give rise to a quantum field theory with zero-modes was studied in detail in Ref. \cite{Munoz}. Concerning these boundary conditions the two most important results from Ref. \cite{Munoz} are:
 \begin{itemize}
 \item There are boundary conditions in the space ${\cal M}_F$ for which $\Delta_U$ has at most one constant zero-mode, the space ${\cal M}_F^{(0)}$.
 \item The space ${\cal M}_F^{(0)}$ can be characterised in terms of the parameters of the unitary matrices that determine the boundary condition as
 \beq
{\cal M}_F^{(0)}\!\!=\{U\in {\cal M}_F\,\, /\, \vert n_1\vert=1,\,\theta=-n_1 \alpha\}.
\eeq
 \end{itemize}
 In this situation, the spectral function that must be used is given by \cite{Munoz}
 \begin{equation}
  f_U^{(0)}(k)=\left.\frac{h_U(k)}{ k^3}\right\vert_{U\in{\cal M}_F^{(0)}}.
 \end{equation}
Plugging $n_1=\pm1,\,\, \theta=-n_1 \alpha$ into the equation above and taking into account Eq.\eqref{fU} we obtain
 \begin{equation}\label{f0}
 f_U^{(0)}(k)=\frac{k \cos (\alpha ) \sin (k L)+\sin (\alpha ) (1-\cos (k L))}{k^2} .
 \end{equation}
 It is of note that the zeros of $ f_U^{(0)}(k)$, i. e. $Z(f^{(0)}_U) $, characterise the non-zero spectrum of the discrete transverse momenta. Therefore the whole spectrum of transverse momenta when $U\in {\cal M}_F^{(0)}$ is $$\sigma(\Delta_U)=\{0\}\cup Z(f^{(0)}_U) .$$
 Hence, in this case the summation over the spectrum $\sigma(\Delta_U)$ splits into two terms,
 \begin{equation}
 \Delta^{(0)}_T{\cal F}={\color{black}A}\int \frac{d^{2} \vec{k_\|}}{{\color{black}(2\pi)^2}} \left[B(k_{\|}, T)+ \!\!\!\!\sum_{k\in Z(f^{(0)}_U) }\!\!\!\! B(\omega, T)\right].
 \end{equation}
 The integration
 \begin{equation}\label{zm-ener}
 {\color{black}A}\int \frac{d^{2} \vec{k_\|}}{{\color{black}(2\pi)^2}} B(k_{\|}, T)={\color{black}A}\int_0^\infty \frac{d k_\|}{{\color{black}2\pi}}k_\| B(k_{\|}, T)
 \end{equation}
 accounts for all the field modes characterised by frequencies $\omega=\sqrt{k_\|^2+0^2}$, that are not included when we perform the summation over the zeroes of $f_U^{(0)}(k)$. The integral \eqref{zm-ener} can be obtained analytically,
 \begin{equation}
{\color{black}A}\int_0^\infty \frac{d k_\|}{{\color{black}2\pi}}k_\| B(k_{\|}, T)=-{\color{black}A}\frac{T^3 \zeta (3)}{2 \pi }.
 \end{equation}
 \noindent Hence the temperature dependent part of the Helmholtz free energy for the case in which there is a zero mode reads,
 \begin{equation}\label{zm-dtf}
 \Delta^{(0)}_T{\cal F}=-\frac{{\color{black}A} \zeta (3)}{2 \pi }T^3+{\color{black}A}\int \frac{d^{2} \vec{k_\|}}{{\color{black}(2\pi)^2}} \sum_{k\in Z(f^{(0)}_U) }\!\! B(\omega, T).
 \end{equation}
  The extra term arising due to the existence of a transverse zero-mode will not contribute to the force between plates since it does not depend on the distance between plates. Nevertheless, it contributes to the dominant term of the free energy at  high temperature since
 \begin{equation}
 -\frac{T^3 \zeta (3)}{2 \pi }\simeq -0.191313 T^3,
 \end{equation}
as discussed in detail in the Appendix.
 The second term in \eqref{zm-dtf} can be again computed using \eqref{eq18} replacing $f_U(z)$ by $f_U^{(0)}(z)$. 
 \begin{figure}[H]
\centering
    \includegraphics[width=.50\textwidth]{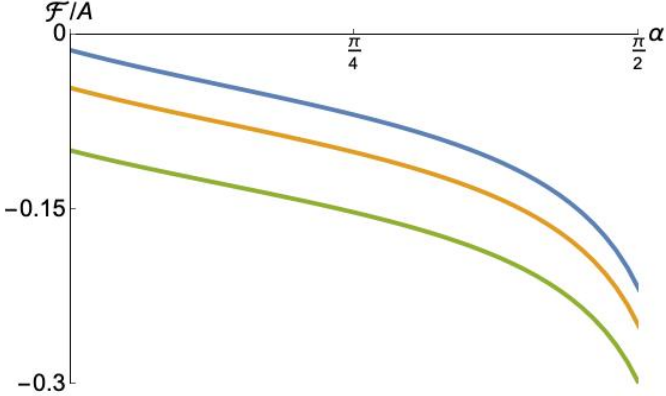}
     \caption{${\cal F}/A$ for $U\in{\cal M}_F^{(0)}$ at low temperatures: $T=0$ (blue line), $T=0.55$ (yellow line), and $T=0.75$ (green line).}
     \label{fig9}
\end{figure}
\begin{figure}[H]
\centering
    \includegraphics[width=0.50\textwidth]{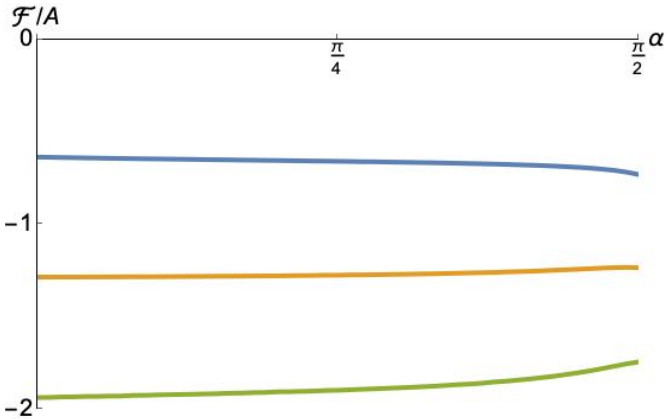}
     \caption{${\cal F}/A$ for $U\in{\cal M}_F^{(0)}$ at high temperatures: $T=1.35$ (blue line), $T=1.65$ (yellow line), and $T=1.85$ (green line).}
     \label{fig10}
\end{figure}
\begin{figure*}
\centering
    \includegraphics[width=1\textwidth]{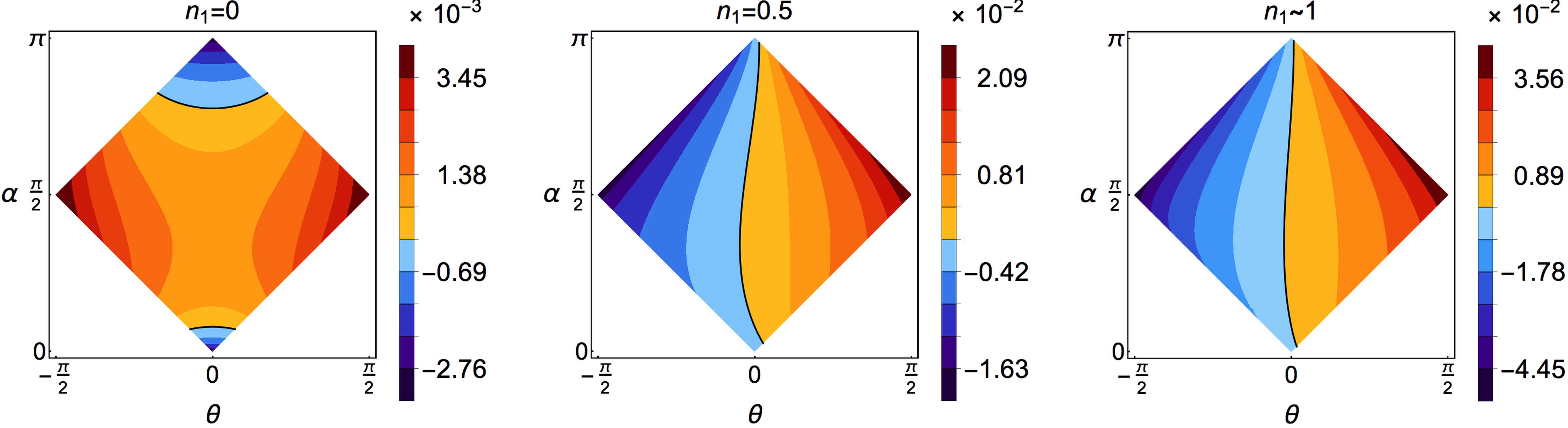}
     \caption{Pressure in the $\alpha-\theta$ plane for different values of $n_1$. In these plots, $T=0.35$ and $L=1$.}
     \label{figp}
\end{figure*}
 In Figs. \ref{fig9} and \ref{fig10} we show the numerical computation of $\left.{\cal F}\right\vert_{{\cal M}_0}$ for different values of the temperature as a function of the free parameter $\alpha$ that determines the elements of ${\cal M}_F^{(0)}$. Periodic boundary condition ($\alpha=\pi/2$) corresponds to the minimum of $\left.{\cal F}\right\vert_{{\cal M}_0}$ for any $T<1$ (see Fig \ref{fig9}). In contrast, when the temperature is sufficiently high (see Fig. \ref{fig10}) $\left.{\cal F}\right\vert_{{\cal M}_0}$ becomes a monotonically increasing function of $\alpha$, so the minimum of $\left.{\cal F}\right\vert_{{\cal M}_0}$ occurs at Neumann boundary condition ($\alpha=0$).

\section{Finite temperature Casimir force. Attraction, repulsion and no-force boundary conditions}\label{sec4}

The finite temperature force per unit area of the plates, i. e. the thermal pressure,  can be obtained from \eqref{e15}. Since only those terms that depend on the distance between plates contribute to the pressure in this case there is no need to make any distinction between boundary conditions in ${\cal M}_F^{(0)}$ and the rest of boundary conditions in ${\cal M}_F$ because the zero mode contribution in \eqref{zm-dtf} does not depend on the distance $L$. Taking this into account the general formula for the temperature dependent part of the quantum vacuum pressure reads,
\begin{eqnarray}\label{eq42}
&&\Delta_T P= \int_0^\infty \frac{d\xi}{2\pi i } \left[I_3(\xi e^{i\gamma},T) \partial_L\partial_\xi \log f_U(\xi e^{i\gamma}) -\right.\nonumber\\
&&\left. \hspace{2cm} I_3(\xi e^{-i\gamma},T) \partial_L\partial_\xi \log f_U(\xi e^{-i\gamma})\right].
\end{eqnarray}

\noindent In addition the zero-temperature pressure can be easily obtained from formula \eqref{e0fin} if we take into account that for any $U\in{\cal M}_F$,
\begin{equation}
E_U^{(3)}=A \frac{c_U^{(3)}}{L^3},
\end{equation}
being $c_U^{(3)}$ a coefficient that does not depend on the distance between plates $L$ \cite{Asorey}. Hence the zero temperature quantum vacuum pressure is
\begin{eqnarray}\label{eq44}
&&P(T=0)=\frac{1}{A}\frac{3 E_U^{(3)}}{L}=\frac{3}{L} \frac{1}{6\pi^2}\times\nonumber\\
&&\int_0^\infty \!\!\!\!\!dk\,k^3\!\!\left[ L-\partial_k\log\left( \frac{h_U(ik,L)}{h_U^\infty(ik)}\right) \right]\!.
\end{eqnarray}
Putting together formulas \eqref{eq42} and \eqref{eq44} we obtain the quantum vacuum pressure for any temperature and any boundary condition $U\in{\cal M}_F$,
\begin{equation}\label{eq45}
P(T)=P(T=0)+\Delta_T P.
\end{equation}
It is of note that since $\Delta_T P$ does not scale with the distance between plates as $L^{-4}$, the regions where the force becomes attractive, repulsive or zero do not match the regions when the total vacuum energy ${\cal F}$ is negative, positive, or zero.
In Fig.\ref{figp} we show the numerical values of the pressure at finite temperature computed using formulas \eqref{eq42}, \eqref{eq44}, and \eqref{eq45}. As can be seen the pressure still gives rise to attraction, repulsion or no-force regimes when the temperature is low enough. In particular, the minimum pressure is obtained for periodic boundary conditions (see plot for $n_1=1$, left-hand-side corner in Fig.\ref{figp}), and the maximum pressure for anti-periodic boundary conditions (see plot for $n_1=1$, right-hand-side corner in Fig.\ref{figp}) , as it is also found for $T=0$ (see Ref. \cite{Asorey}). The strongest temperature effect at low temperature happens when $n_1=0$ where the attractive regime almost disappears with respect the $T=0$ case (see plot $n_1=0$ in Fig.\ref{figp} and its analogue for $T=0$ from Ref. \cite{Asorey}).

\subsection{Critical temperature: the fully repulsive regime}
\begin{figure}
\centering
    \includegraphics[width=.40\textwidth]{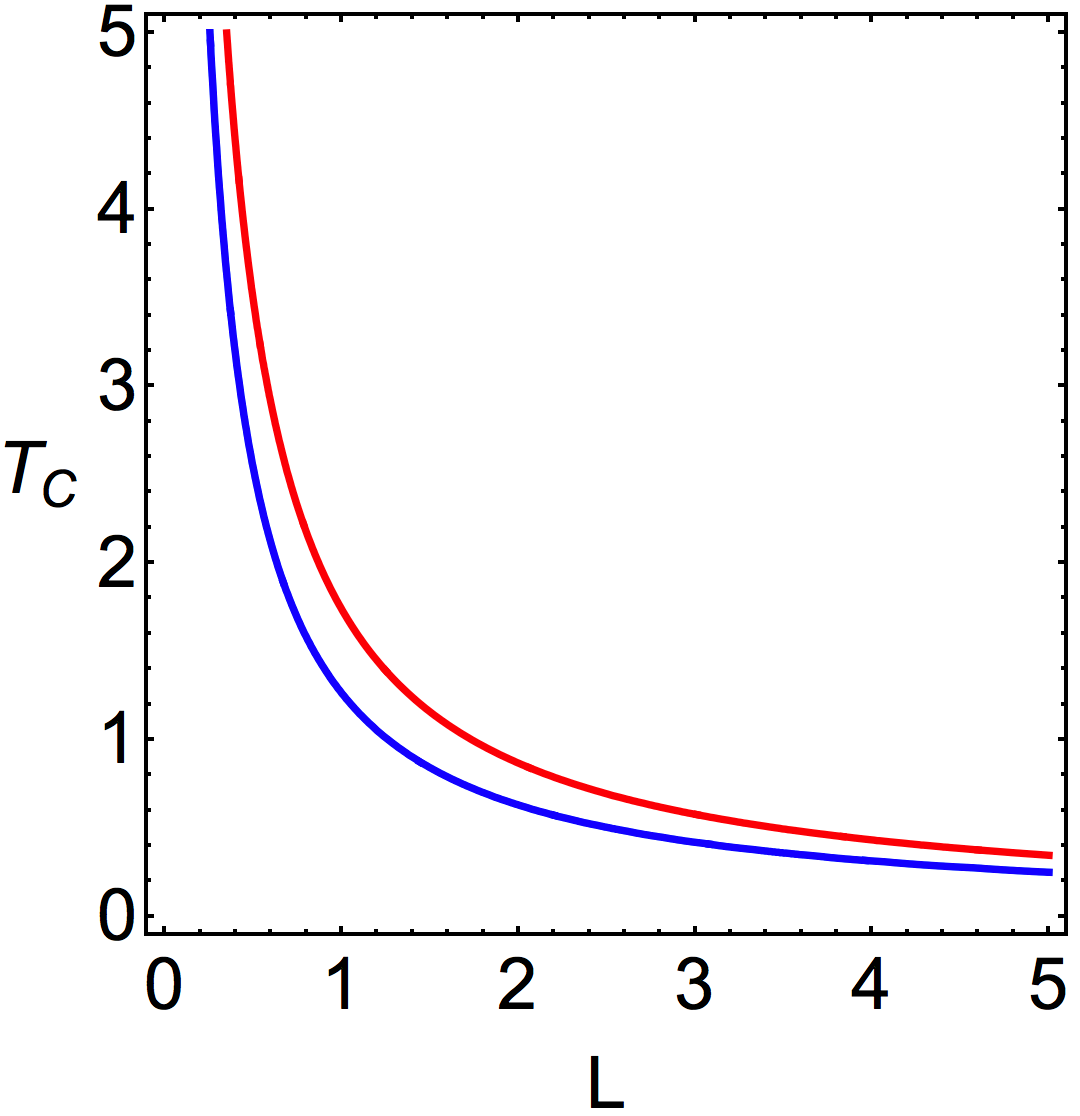}
     \caption{$T_c^{\cal F}$ (blue line) and $T_c^P$ (red line) as functions of the distance between plates $L$.}
     \label{figtc2}
\end{figure}

As it is expected from the previous results, as the temperature increases the quantum vacuum pressure will be dominated by the thermal fluctuations, which tend to produce a repulsive force. Hence the temperature at which the minimum pressure is equal to zero defines a critical temperature $T_c^P$ that separates the thermal-dominated regime (if $T>T_c^P$there is only repulsion due to the thermal fluctuations for any $U\in{\cal M}_F$) and the quantum vacuum fluctuations dominated regime (if $T<T_c^P$there can be attractive, repulsive or null quantum vacuum pressure). 
From Ref. \cite{Asorey}, it is known that the minimum quantum vacuum pressure at zero temperature occurs for periodic boundary conditions,
\begin{equation*}
U_p=\left(
\begin{array}{cc}
 0 & 1 \\
 1 & 0 \\
\end{array}
\right).
\end{equation*}
Taking this into account, the equation for the critical temperature $T_c^P$ at a given distance between plates, $L$, is $P(U_p,T_c^P)=0$, from which
\begin{equation}\label{tcp}
\Delta_T P(U_p,T_c^P)=-P(U_p,T=0),
\end{equation} 
where $P(U_p,T=0)$ is given by \cite{Asorey},
\begin{equation*}
P(U_p,T=0)=-\frac{\pi^2}{15 L^4}.
\end{equation*}
Eq. \eqref{tcp} can not be solved analytically, so we have to proceed by using numerical methods. In Fig. \ref{figtc2} we show the numerical results for $T_c^P$ and $T_c^{\cal F}$ together. As can be seen the temperature at which the possibility of having attractive quantum vacuum pressure disappears ($T_c^P$) is higher than the critical temperature ($T_c^{\cal F}$) at which the total quantum vacuum energy ${\cal F}$ becomes definite negative.

\section{Conclusions and further comments }\label{sec5}
In this paper we have considered a massless scalar field confined between two plane, isotropic and homogeneous parallel plates mimicked by sufficiently general boundary conditions compatible with the principles of the theory of quantum fields at finite temperature. We have computed and analysed the Helmholtz free energy, the entropy and the pressure as functions of the temperature and the free parameters entering the boundary condition.

Concerning the Helmholtz free energy, the main result obtained is the possibility of having a change in its sign for temperatures under certain critical $T_c^{\cal F}$ (see Fig. \ref{fig6}, and Figs. \ref{fig2}-\ref{fig4}). As happens at zero temperature (see Ref. \cite{Asorey}), at finite temperature and $T<T_{C}^{\mathcal{F}}$ the maximum of the free energy occurs for anti-periodic boundary conditions, and the minimum is reached for periodic boundary conditions (see Fig. \ref{fig3}). Nevertheless, since the finite temperature correction to the free energy $\Delta_T{\cal F}$ has its minimum at Neumann boundary conditions\footnote{Neumann boundary condition corresponds to $\theta=\alpha=0$.} (see Figs. \ref{fig2}-\ref{fig5},  \ref{fig9} and \ref{fig10}) and its maximum at Dirichlet boundary conditions\footnote{Dirichlet boundary condition corresponds to $\theta=0$ and $\alpha=\pi$.}, when the temperature is high enough the maximum and the minimum of ${\cal F}$ take place at Dirichlet and Neumann boundary conditions, respectively.

Regarding the entropy as a function of the temperature and the free parameters mimicking the plates, we have found that the one-loop quantum correction to the entropy is positive definite for any temperature and any $U\in {\cal M}_F$. In relation with previous works where there have been found negative one-loop quantum corrections to the entropy in Casimir-like systems \cite{geye05-72-022111,bord18-98-085010,bord18-51-455001,milt17-96-085007,li16-94-085010}, suggesting certain instabilities of the quantum system \cite{thir70-235-339}, we conclude that the quantum system of a scalar field confined between two plates mimicked by the sufficiently general boundary conditions is always thermodynamically stable. Moreover, for any temperature the maximum entropy is reached for Neumann boundary conditions, while the minimum is obtained for Dirichlet boundary conditions (see Figs. \ref{fig7} and \ref{fig8}). In Ref. \cite{asojmc} it was computed how the renormalization group transformations for massless scalar fields transform the boundary condition defined by a unitary matrix $U\in{\cal M}_F$, i. e., the boundary renormalization group transformations. It is remarkable that the extrema of the entropy for any temperature occur for the most unstable (Dirichlet boundary condition) and stable (Neumann boundary condition) boundary renormalization group flow fixed points which were studied in Ref. \cite{asojmc}. In addition taking into account the results presented in Ref. \cite{bor-epjc} we can ensure that when there exists a potential with compact support between plates, the entropy in general can become negative for certain potentials, making the quantum system thermodynamically more unstable than its classical analogue (see Ref. \cite{thir70-235-339})

\begin{figure}
\centering
    \includegraphics[width=.47\textwidth]{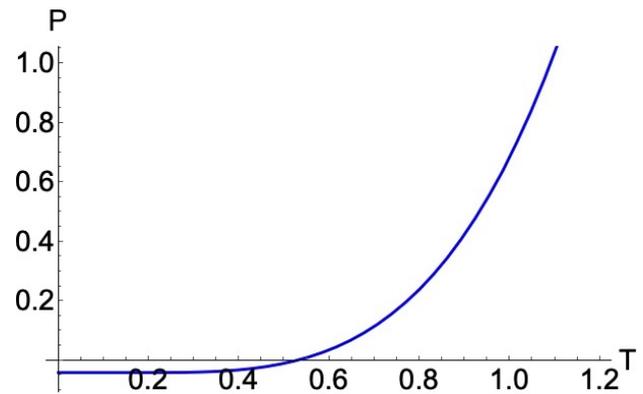}
     \caption{Quantum vacuum pressure as a function of the temperature for Neumann/Dirichlet boundary conditions and for $L=1$.}
     \label{fig13}
\end{figure}

For the Casimir pressure at finite temperature we have obtained the critical temperature $T_c^P$ that separates the thermal fluctuation dominated regime ($T>T_c^P$), and the zero-temeperature fluctuation dominated regime  ($T<T_c^P$) (see Fig. \ref{figtc2}). On the one hand, when $T<T_c^P$ the quantum vacuum fluctuations at zero temperature still dominate the pressure behaviour of the system. Hence, there exist boundary conditions that produce attractive, repulsive or null force between plates (see Fig. \ref{figp}).  On the other hand, for $T>T_c^P$ the thermal fluctuations become dominant giving rise to a repulsive force between plates for any boundary condition $U\in {\cal M}_F$. In addition, as it happens at zero temperature, when $T<T_c^P$ the maximum and minimum of the quantum vacuum force occur for anti-periodic and periodic boundary conditions, respectively.  As a consequence, we can conclude that the theorem of Kenneth and Klich that states\footnote{The main result of the paper is that Casimir-like systems with mirror symmetry produce attractive Casimir forces.} the {\it opposites attract} \cite{Kenneth,kenneth-prb08} only holds for $T<T_c^P$. For instance, it can be seen in Fig. \ref{fig13} that the sign of the Casimir pressure for Dirichlet and Neumann boundary conditions varies with the temperature.

It is of note that our results can be generalised to arbitrary dimension. If the physical space with boundary in which the scalar field is confined is given by $[0,L]\times\mathbb{R}^D$, then all our arguments and formulas to compute the free energy, the entropy and the pressure remain valid by just replacing $I_3$ defined in Eq. \eqref{i3-def} with
\begin{equation}
I_D\equiv T\int_{\mathbb{R}^{2}} \frac{d^{D} \vec{k_\|}}{{\color{black}(2\pi)^D}}\,   \log \left( 1-e^{-\frac{\sqrt{\vec{k_\|}^2+k^2}}{ T}}\right).
\end{equation}
This integral can be computed in terms of more complicated combinations of polylogarithms of higher order. Nevertheless since in any case the arguments of the polylogarithms will be $e^{-k/T}$ it is ensured that non of them will go through the branch cut when performing the integration in $k$ to sum over the orthogonal modes. For the boundary conditions $U\in{\cal M}_F^{(0)}$ the dominant contribution of the zero-mode to the free energy will be given by 
\begin{equation}
\Delta_T^{(0)}{\cal F}_{\rm zm}=\frac{A T^D\Gamma\left( \frac{1+D}{2}\right)}{\pi^{\frac{1+D}{2}}}\zeta(1+D)
\end{equation}

\appendix

\section{High and low temperature expansions}
The analytic formulas used to compute the numerical results throughout the paper are valid for any temperature. Nevertheless, we can obtain some simplified expressions in the low and high temperature limits. In the first place, it is customary  in the literature to define these limits in terms of the quantity $TL$. That is, $TL\ll1$ corresponds to the low temperature limit whereas $TL\gg1$ does to the high temperature limit.
Both limits have been well studied in the literature for the most common boundary conditions \cite{Bordagbook,Klaus}, where the length scale $a$ appearing in Eq \eqref{fU} is irrelevant. Our purpose in this appendix is to extend the definition of these limits in order to account for the general boundary conditions studied in this paper. It will be shown that the scale $a$ becomes relevant at the low temperature regime.

\subsection{The low temperature expansion}
For the most common boundary conditions, i.e., Dirichlet, Neumann and (anti)periodic, the low temperature approximation to the Helmholtz free energy is computed in terms of the lowest frequency of the field modes \cite{Bordagbook,Klaus,Miltonbook}. It is very standard to write down the dominant contribution to $\Delta_T{\cal F}$ in the low temperature regime when there are no zero-modes  as
\begin{equation}\label{lambda0}
\Delta_T\mathcal{F}\simeq-\omega_{0}\frac{A}{2\pi} T^2 \exp\left[-\frac{\omega_{0}}{T}\right],\quad LT\ll1,
\end{equation}
with $\omega_0$ being the lowest field mode. Likewise, as for the case that there are zero-modes the dominant contribution is the first term on the lef-hand-side in Eq. \eqref{zm-dtf}, i. e. 
\begin{equation}\label{FE_zeromodes}
\Delta_T^{(0)}{\cal F}_{\rm zm} = -\frac{A \zeta_R(3)}{2\pi} T^{3},\quad LT\ll1.
\end{equation} 
In the following we will show that the low temperature expansion \eqref{lambda0} is not valid when we deal with boundary conditions given by unitary operators $U\in{\cal M}_F-{\cal M}_F^{(0)}$ that are very close to ${\cal M}_F^{(0)}$. In addition, we will give a low temperature expansion for these boundary conditions and the corresponding free energy. At the same time, we will show that a more general definition of the low temperature limit is needed.

In previous sections, for the sake of simplicity, the length scale $a$ has been set to one. In this section, we set $a$ free to vary and study the analytical properties of $\Delta_{T}\mathcal{F}$ as a function of the boundary condition parameters, making special emphasis on the relationship of those parameters, $\alpha$, $\theta$ and $n_{1}$, with the fundamental length scales of the system, $L$ and $a$ at low temperature $T$.
We start writing for any $U\in{\cal M}_F$,
\begin{equation}\label{gendtf}
 \Delta_T{\cal F}=q_0+A\int \frac{d^{2} \vec{k_\|}}{(2\pi)^2} \sum_{k\in Z(f^{(J)}_U) }\!\! B(\omega, T),
 \end{equation}
where
 \begin{equation}
 q_0=\begin{cases}
 0 & U\in{\cal M}_F-{\cal M}_F^{(0)} \\
 -\frac{A \zeta (3)}{2 \pi }T^3 &{\cal M}_F^{(0)}
\end{cases},
 \end{equation}
 and 
 \begin{equation}
f^{(J)}_U=\begin{cases}
 f_U & U\in{\cal M}_F-{\cal M}_F^{(0)} \\
f^{(0)}_U &U\in{\cal M}_F^{(0)}
\end{cases}.
 \end{equation}

In order to perform the low temperature expansion in the surroundings of $\mathcal{M}_{0}$ we need to develop an approximate solution for the spectrum of normal modes $k_n$ for those boundary conditions that do not have zero modes but that are {\it very close} to those that do.
%
%

\subsubsection{Analytical behaviour of the spectrum around $\mathcal{M}_F^{(0)}$}

In order to provide an analytical result, we will restrict ourselves to variations in a neighbourhood of 
$\mathcal{M}_F^{(0)}$. To do so, we compute the eigenvalues of the spectrum of normal modes in a neighbourhood of $\mathcal{M}_{0}$. That is given by Eq. \eqref{f0}, and the equation that characterises the spectrum of the transverse momenta for the associated selfadjoint extensions can be rewritten as
\begin{equation}
\frac{\cos{\theta} \sin{(kL)}}{k a} - \frac{\sin{\theta}(1-\cos{(k L)})}{(k a)^2} = 0.
\end{equation}
This equation is fulfilled whether
\begin{equation}\label{kn0_solution}
k_{2n}^{(0)} = \frac{2\pi n}{L} \quad \textrm{or} \quad  k_{2n+1}^{(0)} = \frac{\tan{\theta}\tan{(k_{2n-1}^{(0)} L /2)}}{a},
\end{equation}
for $n = 0,1,2,3 \dots$. Notice that, generally, the distance between consecutive eigenvalues depends strongly on the value of $\theta$. In fact, the distance between two consecutive eigenvalues reaches its maximum is for Neumann's boundary condition 
$$\theta=0\Rightarrow \,\,k^{(0)}_{n+1}-k_{n}^{(0)}=\frac{\pi}{L},$$ 
and the minimum distance occurs for periodic boundary conditions ($\theta=-\pi/2$) when the solutions of the equation for $k_{2n-1}^{(0)}$ in Eq. \eqref{kn0_solution} coincide with $k_{2n}^{(0)}=2 \pi n/L$, giving rise to a spectrum with degeneracy two. Although no closed formulas can be provided for the odd modes at generic values of $\theta$, this behaviour guarantees that there is no level crossing in ${\cal M}_F^{(0)}$.


To study the low temperature approximation of $\Delta_T{\cal F}$ for selfadjoint extensions close to ${\cal M}_F^{(0)}$, we must study some analytical properties of the spectrum of such selfadjoint extensions. To do so we will consider the subset of selfadjoint extensions characterised by 
\begin{equation}\label{ueps}
\lbrace U(\alpha,\theta, \vec{ n})\in {\cal M}_F\,\,\text{such that}\,\, \alpha+ \theta = \epsilon, n_1 = 1\rbrace,
\end{equation} 
where $\epsilon > 0$ is a small displacement in the $\alpha\text{-}\theta$ plane. To perform a perturbative study of the spectrum of the selfadjoint extensions in\eqref{ueps} we make the substitutions $\alpha = -\theta + \epsilon$ and $k_n = k_n^{(0)} + \epsilon \delta_n$ in the spectral function $f_U$ given by Eq. \eqref{fug}, and expand it up to first order in $\epsilon$. Solving for $f_U = 0$, we obtain for the lowest transverse mode ($k_0^{(0)}=0$),
\begin{equation}\label{k_0_epsilon}
k_0 = \sqrt{\frac{\epsilon}{aL}}.
\end{equation}
It is of note that, in the limit $\epsilon \rightarrow 0$, $k_0$ goes smoothly to zero as expected.
Moreover the condition for the selfadjoint extension $\Delta_U\in{\cal M}_F$ to be close to ${\cal M}_F^{(0)}$ can be re-formulated as the requirement $k_0 L\ll1$ i. e.,
\begin{equation}
\epsilon\ll a/L.
\end{equation}
 Thus, the spectrum varies continuously as a function of the boundary condition parameter $\theta$ in a  neighbourhood of $\mathcal{M}_F^{(0)}$, regardless of the length scales. As for the modes with $n\geq 1$ we get
\begin{equation}
k_n = k_n^{(0)} + \delta_n \epsilon,
\end{equation}
where
\begin{equation}
\delta_n = \frac{2 k_n^{(0)} a   - ((k_n^{(0)} a)^2 + 1) \tan{\theta}\tan{(k_n^{(0)} L)}}{2 k_n^{(0)} a (k_n^{(0)} a L  + \tan{(k_n^{(0)} L)}(a - L\tan{\theta}))}.
\end{equation}
 In general, $\delta_n$ depends on the parameter $\theta$ as well as on $L$, $a$ and $n_1$ through $k_n^{(0)}$.   

\subsubsection{Analytical behaviour of the thermal corrections to the free energy at low-temperature}

We finalise this section studying the analyticity of the free energy as a function of the boundary parameters in a  neighbourhood of $\mathcal{M}_F^{(0)}$. To this end, we compute the thermal corrections to the free energy as a function of the dimensionless parameter 
\begin{equation}
\chi\equiv\epsilon/LaT^{2},
\end{equation}
in the low-temperature limit, $LT\ll1$. 

As explained previously, for $LT\ll1$, all along $\mathcal{M}_F^{(0)}$ the free energy in Eq.(\ref{FE_zeromodes}) is dominated by the contribution of the zero mode $k_{0}=0$ given by Eq. \eqref{FE_zeromodes}. In the neighbourhood of $\mathcal{M}_F^{(0)}$ given by the set of parameters $\lbrace \alpha= - \theta + \epsilon, \theta ,n_1 = 1\rbrace$,  the lowest mode $k_{0}=\sqrt{\epsilon/aL}>0$ is still the dominant contribution to $\Delta_{T}\mathcal{F}$. However,  two asymptotic behaviours can be distinguished in $\Delta_{T}\mathcal{F}$ as a function of $k_{0}/T$. In the first case, when $$k_{0}/T\ll1\Rightarrow \chi=\epsilon/aLT^{2}\ll1,$$ $I_3$ in Eq.(26) can be expanded around the zero mode up to leading order in $k^{2}_{0}/T^{2}\ll1$. That yields a logarithmic correction to  $\Delta_T^{(0)}{\cal F}_{\rm zm}$,
\begin{equation}
\left.\Delta_{T}\mathcal{F}\right\vert_{\chi\ll1}\simeq\Delta_T^{(0)}{\cal F}_{\rm zm}-\frac{1}{8\pi}\frac{A}{aL}T\epsilon\log{(\epsilon/aLT^{2})},\label{eqDF1}
\end{equation}
On the contrary, for $$k_{0}/T\gtrsim1\Rightarrow \chi=\epsilon/aLT^{2}>1,$$
\begin{figure*}
\centering
    \includegraphics[width=1\textwidth]{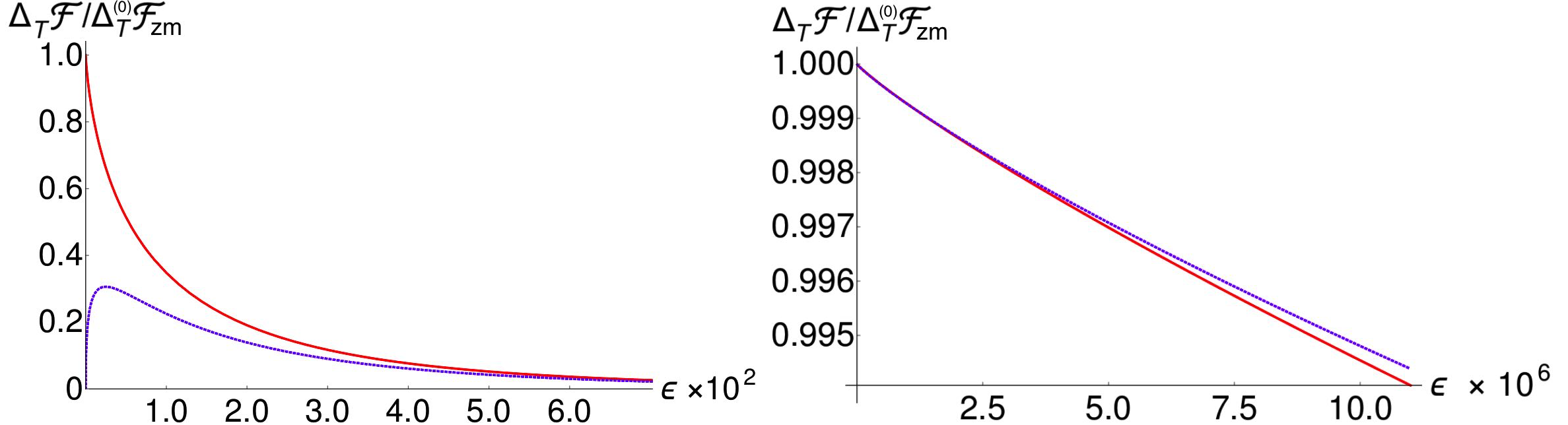}
     \caption{Graphical representation of $\Delta_{T}\mathcal{F}/\Delta_T^{(0)}{\cal F}_{\rm zm}$ as a function of the parameter $\epsilon$ in the subset $\{\alpha=-\theta+\epsilon,\theta,n_{1}=1\}$, as given by Eq.\eqref{eq18} (solid line in red), and their asymptotic approximations of Eq.\eqref{eqDF1} (dashed line in blue, RIGHT) and Eq.\eqref{eqDF2} (dashed line in blue, LEFT). The numerical values are $T=1$, $L=0.1$, and $a=0.025$.}
     \label{figb}
\end{figure*}
the Boltzmann factor happens to be exponentially suppressed, and the contribution of $k_{0}$ is the one given by Eq. \eqref{lambda0} as shown in standard references (c.f. \cite{Bordagbook})
\begin{equation}
\left.\Delta_{T}\mathcal{F}\right\vert_{\chi>1}\simeq-\frac{A\sqrt{\epsilon}}{2\pi\sqrt{aL}}T^{2}\exp\left[-\sqrt{\frac{\epsilon}{LaT^2}}\right].\label{eqDF2}
\end{equation}
Therefore, we conclude that the low  temperature limit in a neighbourhood of $\mathcal{M}_{0}$ needs to be refined. That is, on top of the customary criterion $LT\ll1$, an additional condition upon $k_{0}/T$ must be considered. In Fig.\ref{figb} we compare $\left.\Delta_{T}\mathcal{F}\right\vert_{\chi\ll1}$ and $\left.\Delta_{T}\mathcal{F}\right\vert_{\chi>1}$ from Eqs. \eqref{eqDF1} and \eqref{eqDF2} respectively, and compare both approximations with $\Delta_{T}\mathcal{F}$ using the exact formula \eqref{eq18} for boundary conditions of the form $$U\left( \alpha= - \theta + \epsilon, \theta ,n_1 = 1\right)\in \mathcal{M}_F-\mathcal{M}^{(0)}_F,$$
close enough to ${\cal M}_F^{(0)}$. It can be seen that, for a fixed values of $T$ and $L$ such that $TL\ll1$, when $\epsilon$ is sufficiently small such that $k_{0}/T\ll1$, the low temperature approximation \eqref{eqDF1} is much better than the one obtained from \eqref{eqDF2}. Therefore, we conclude that the standard low temperature approximation given by Eq. \eqref{eqDF2} is valid only when the boundary conditions are not close to ${\cal M}_F^{(0)}$ for a given temperature, i.e., when $k_{0}/T>1$.

\subsection{The high temperature limit}
The high temperature expansion of the Helmholtz free energy can be obtained in terms of the high energy part of the one-particle states spectrum. Following Refs. \cite{Bordagbook,Klaus} the latter can be written using zeta function regularization as\footnote{The physical limit is obtained taking $s\to 0$.}
\begin{equation}\label{fzeta}
{\cal F}(s)=q_0-{1\over 2}\frac{\partial}{\partial s}\mu^{2s}\int_0^\infty\frac{dt}{t}\frac{t^s}{\Gamma(s)}K_T(t)K^{(3)}_{U}(t),
\end{equation}
being
\begin{equation}
K_T(t)=T+2T\sum_{\ell=1}^\infty e^{-t\xi_\ell^2};\,\,\xi_\ell=2\pi T\ell,
\end{equation}
and
\begin{equation}\label{a17}
K^{(3)}_U(t)=\sum_{\omega\in\sigma(\Delta_U)} e^{-t\omega}.
\end{equation}
the heat trace for the selfadjoint extension $-\Delta_U$. After subtracting the divergences in \eqref{fzeta} the series expansion for high temperature, i. e. $TL\to\infty$ can be written down in terms of heat kernel coefficients associated to $-\Delta_U$. Remember that
\begin{equation}
\Delta_U=\Delta_{\mathbb{R}^2}+\Delta_{[0,L]}^U,\quad[\Delta_{\mathbb{R}^2},\Delta_{[0,L]}^U]=0
\end{equation}
being $\Delta_{[0,L]}^U$ the selfadjoint extension of the operator $d^2/dx^2$ over the interval $[0,L]$ associated to the boundary condition defined by $U\in{\cal M}_F$. Hence
the heat trace for $\Delta_U$ in Eq. \eqref{a17} factorises as
\begin{equation}\label{a19}
K(t)=K^{(2)}_{\|}(t) K^{(1)}_U(t),
\end{equation}
being $K^{(2)}_\|(t)$ and $K^{(1)}_U(t)$ the heat traces for $-\Delta_{\mathbb{R}^2}$ and $-\Delta_{[0,L]}^U$ respectively.
From Eq. \eqref{a19} it is obvious that the heat kernel coefficients of $-\Delta_U$ can be written in terms of products of heat kernel coefficients for $-\Delta_{\mathbb{R}^2}$ and heat kernel coefficients of $-\Delta_{[0,L]}^U$. Taking into account that the heat kernel coefficients for $-\Delta_{\mathbb{R}^2}$ can be found in standard books (see e. g. Ref. \cite{Klaus}) and that the heat kernel coefficients for the operator $-\Delta_{[0,L]}^U$ were recently computed as functions of the matrix $U\in{\cal M}_F$ in Ref. \cite{Munoz}, the high temperature expansion is fully determined and does not require any extra attention.

\begin{acknowledgements}
The authors are grateful to the Spanish Government-MINECO (MTM2014- 57129-C2-1-P) and the {\it Junta de Castilla y Le\'on} (BU229P18, VA137G18 and VA057U16)   for the financial support received.  LSS is grateful to the Spanish Government-MINECO for the FPU-fellowships programme (FPU18/00957). MTF acknowledges financial support from the European Social Fund, the Operational Programme of {\it Junta de Castilla y Le\'on} and the regional Ministry of Education. JMMC acknowledges the fruitful discussions with M. Bordag, K. Kirsten, G. Fucci, M. Asorey and I. Pirozhenko.
\end{acknowledgements}




\end{document}